\documentclass[runningheads]{llncs}
\usepackage{graphicx}
\usepackage{booktabs}
\usepackage[colorinlistoftodos]{todonotes}
\usepackage{pifont}

\usepackage[symbol]{footmisc}

\usepackage{enumitem,amssymb}
\newlist{todolist}{itemize}{2}
\setlist[todolist]{label=$\square$}

\begin{document}
\title{Graph Convolutional Neural Networks for Automated Echocardiography View Recognition: A Holistic Approach}

\titlerunning{GCN for pose estimation and view recognition}
%\author{%

% UIO
\author{Sarina Thomas \inst{1} \and
Cristiana Tiago \inst{2} \and
Børge Solli Andreassen \inst{1} \and
Svein Arne Aase \inst{2} \and
Jurica Šprem\inst{2} \and
Erik Steen \inst{2} \and
Anne Solberg\inst{1} \and
Guy Ben-Yosef \inst{3}}

\authorrunning{S. Thomas et al.}
\institute{
Division of Digital Signal and image processing, University of Oslo, Oslo, Norway \and
GE Vingmed Ultrasound, GE Healthcare, Horten, Norway \and
GE Research, Niskayuna, New York, USA \\ \email{sarinat@uio.no}}

%Acknowlegdement
% \author[2]{Daria \snm{Kulikova}}
%\author[2]{Anna \snm{Novikova}}
\authorrunning{Anonymous}
%\institute{Anonymous organization\\
%\email{****@****.***}\\

\maketitle              
% typeset the header of the contribution
%
\begin{abstract}

To facilitate diagnosis on cardiac ultrasound (US), clinical practice has established several standard views of the heart, which serve as reference points for diagnostic measurements and define viewports from which images are acquired. Automatic view recognition involves grouping those images into classes of standard views. Although deep learning techniques have been successful in achieving this, they still struggle with fully verifying the suitability of an image for specific measurements due to factors like the correct location, pose, and potential occlusions of cardiac structures. Our approach goes beyond view classification and incorporates a 3D mesh reconstruction of the heart that enables several more downstream tasks, like segmentation and pose estimation. In this work, we explore learning 3D heart meshes via graph convolutions, using similar techniques to learn 3D meshes in natural images, such as human pose estimation. As the availability of fully annotated 3D images is limited, we generate synthetic US images from 3D meshes by training an adversarial denoising diffusion model. Experiments were conducted on synthetic and clinical cases for view recognition and structure detection. The approach yielded good performance on synthetic images and, despite being exclusively trained on synthetic data, it already showed potential when applied to clinical images. With this proof-of-concept, we aim to demonstrate the benefits of graphs to improve cardiac view recognition that can ultimately lead to better efficiency in cardiac diagnosis. 

\end{abstract}

\section{Introduction}
\label{sec:Introduction}
In the field of diagnosing heart diseases, cardiovascular ultrasound --- also known as echocardiography --- is the most commonly used technique due to its accessibility, instantaneous results, and lack of ionizing radiation. Cardiac diagnosis includes the collection of multiple ultrasound (US) images of the heart to identify possible pathologies. In these images, measurements of anatomical structures are obtained to identify pathological deviations from standard norms. The accuracy of measurements is heavily dependent on the quality of the image. The American Society of Echocardiography (ASE) guidelines \cite{ASE_guidelines} recommend obtaining specific standardized views to ensure quality. These views ideally show all the required anatomical landmarks. However, anatomical variations, for example, the body mass index, can complicate the obtaining of correct views, leading to multiple attempts and non-standard views. Diagnostic measurements are time-consuming as clinical experts must carefully identify suitable images. One solution to speed up this process is to employ machine learning methods to automatically select appropriate views, called automatic view recognition. State-of-the-art methods utilize deep learning techniques, which involve training convolutional neural networks (CNNs) on large datasets to directly classify view labels. However, %it's important to note that
an automated selection may not result in a suitable image. Several factors come into play, such as whether structures of interest are fully visible. These factors vary for each view and measurement following the guidelines \cite{ASE_guidelines}. We propose a holistic measurement-centric approach to assess the suitability of images for diagnosis beyond selecting the correct view. Our approach addresses multiple tasks relevant to echocardiography quality control: By estimating the 3D heart pose in relation to the US probe, it becomes possible to select the correct view, localize chambers, and identify any issues such as sector intersections, misalignment, or occlusions, consequently aiding in discarding unsuitable images. Our contribution includes: (1) a \textbf{graph convolutional neural network for pose regression} of the US plane w.r.t. a 3D mesh, (2) a multi-structure graph for \textbf{localizing all chambers and ventricles}, and (3) exploring a \textbf{diffusion-based approach} to overcome the lack of 3D annotations. To our knowledge, this is the first work to apply a graph approach to 2D-3D reconstruction in echocardiography, demonstrating simultaneous localization of chambers and view prediction.

\section{Related work} 
\label{sec:related_work}
Although our work is geared toward automatic view recognition, it touches on a variety of research fields. Therefore, this chapter reviews view recognition approaches but also the current state-of-the-art in graph neural networks (GNN) for medical image segmentation and 2D-3D reconstruction. In the literature, automatic view recognition is mainly regarded a classification problem that predicts a view label defined by the ASE. Various CNN architectures have been explored, including different variations based on ResNet~\cite{10.3389/fped.2021.770182}, InceptionNet~\cite{OSTVIK2019374} and VGG~\cite{doi:10.1161/CIRCULATIONAHA.118.034338}. The studies vary in the number of training samples and view classes. More recent approaches also include contrastive learning and incorporate contrast-enhanced US images~\cite{10.1007/978-3-031-16440-8_33}. All aforementioned approaches have in common that they only output a label without indicating how close the view is in relation to the corresponding standard view. Paseloup et al. \cite{PASDELOUP2023333} estimate probe rotation and tilt by formulating view recognition as a regression problem. Their approach provides the relative probe rotation along with the view; however, it does not include the probe translation.
%A recent study \cite{} combined view classification directly with segmentation of structures but showed it on
% similar idea but shown on very diverse datasets
% https://www.ncbi.nlm.nih.gov/pmc/articles/PMC9310146/#R40
%Real-time echocardiography image analysis and quantification of cardiac indices \cite{ZAMZMI2022102438}
% https://link-springer-com.ezproxy.uio.no/chapter/10.1007/978-3-030-87583-1_15
%https://link-springer-com.ezproxy.uio.no/chapter/10.1007/978-3-031-16440-8_33 
%Seeland MMäder PMulti-view classification with convolutional neural networksPLoS ONE20211610.1371/journal.pone.0245230
%\todo{add head face hand pose estimation related work? Must be fairly short because we don't have space}
Due to their strong ability to model spatial relations, GNNs have been successfully applied to various tasks in medical imaging, demonstrating their potential in segmenting structures in X-Rays \cite{10.1007/978-3-030-87193-2_57} and in spatio-temporal US~\cite{gcn2022miccai}. Kong et al.~\cite{KONG2021102222} were the first to propose a GNN approach to create a 3D representation of the entire heart given a 3D CT or MR scan. Stojanovski et al.~\cite{10.1007/978-3-031-16902-1_9} use a Pix2Vox++ model to predict a 3D voxel representation from multiple 2D US images but without aligning the images to the resulting volume. Our research is founded on the notion of 3D mesh models, which have been derived from natural images such as a human body \cite{Cv2019Grpah}, face~\cite{Lin_2020_CVPR} and hand mesh~\cite{li2022interacting}, and more recently, X-Ray images~\cite{nakao}. In the proposed work, dense graph convolutions are applied in various training methods together with 3D parametric models. Our aim is to expand existing methods and integrate them with generative models~\cite{DBLP:journals/corr/abs-2006-11239} to create synthetic training data.

\begin{figure}[t]
    \centering
    \includegraphics[width=0.95\textwidth]{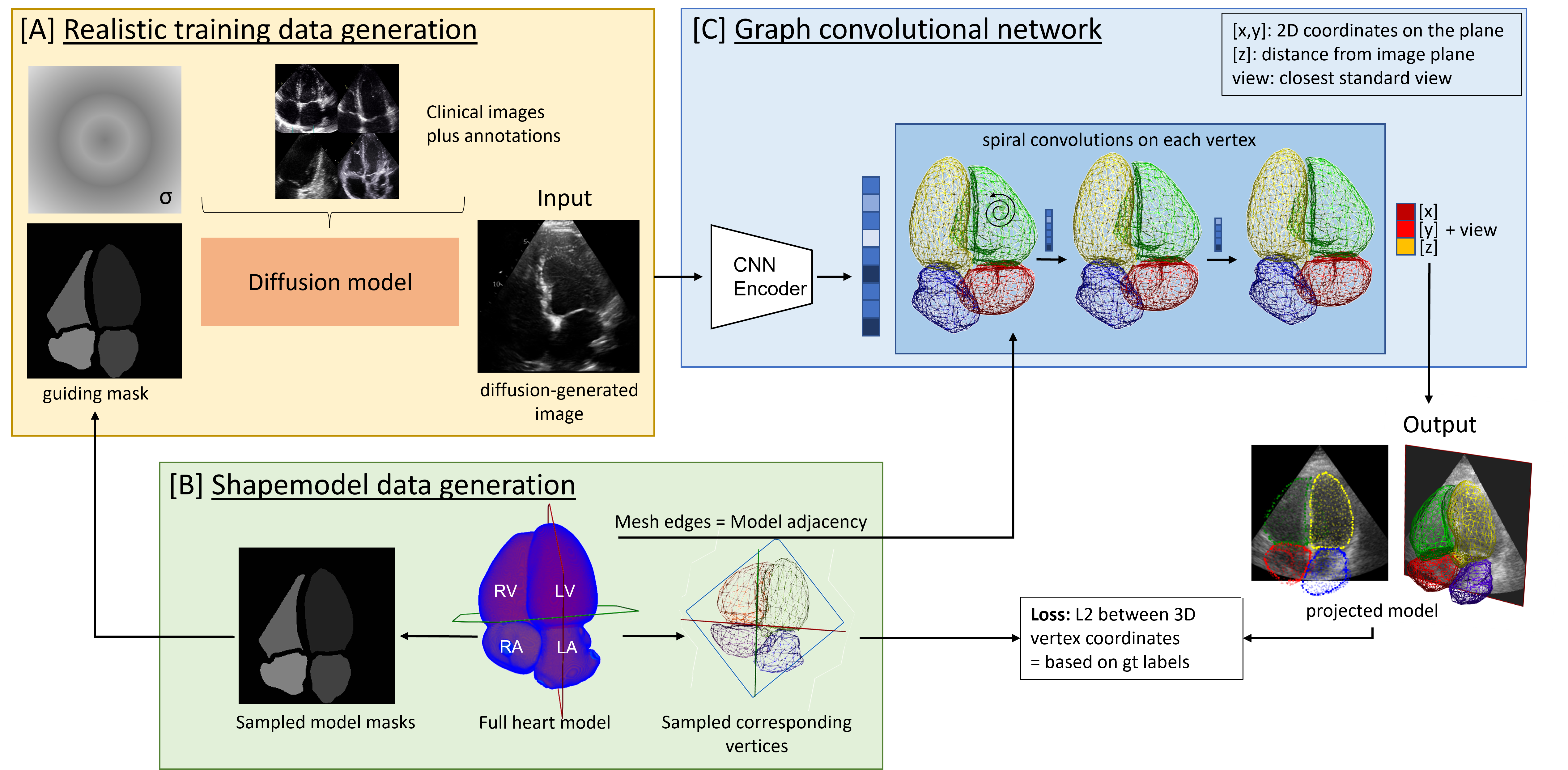}
    \caption{Pipeline overview: A) A diffusion model is trained with real US images and segmentations to generate synthetic images guided by the segmentations of the 3D mesh. B) Synthetic segmentations are sampled from 3D heart meshes along with the 3D coordinates of the model and the view. C) A GCN uses the 3D mesh vertices as nodes and predicts the 3D vertex positions on the 2D image.}
    \label{fig:overview}
\end{figure}

\section{Method}
\label{sec:method}
The proposed pipeline aims to provide additional information on whether an image is suitable for diagnostic measurement. Instead of simply classifying the view label, we want to answer whether 1) relevant structures exist in the image, 2) those structures are fully visible, and 3) the image is taken from a view suitable for the measurement. All these questions can be answered by predicting the 3D position of a cutplane within a virtual shape model. The core of the method, illustrated in Fig. \ref{fig:overview}C, consists of a graph neural network that receives a single US image as input and predicts 3D coordinates of a 4-chamber heart model aligned with this US plane. Training the network in a supervised manner requires images with corresponding 3D coordinates. The following sections introduce the 3D model definition and training data generation. For this work, no annotated clinical 3D data was available. Therefore, a diffusion model was used to create synthetic US images given a guiding segmentation provided by the 3D mesh. %The GCN was trained using resulting images and their corresponding coordinates. %To address the difference between real clinical images and the diffusion-generated images, additional loss functions were introduced.

\subsection{Model embedding}
\label{subsec:methods:modelembedding}
%- original meshes registered 
%- sample closest point from each downsampled point 
The backbone of the pipeline is a 3D model (Fig. \ref{fig:overview}B) that can be aligned to the image. An anatomical structure can be represented as a closed mesh i.e. generated from a binary voxel representation. A mesh comprises 3D vertices and connecting edges. In this study, publically available patient heart meshes from \cite{rodero_cristobal_2021_4590294} with multiple structures are utilized. Four structures are extracted, namely, left ventricle (LV), right ventricle (RV), left atrium (LA) and right atrium (RA). Fine-grained meshes are already registered, but their vertices do not correspond. One important feature in the model design is estimating the correspondences to build a relation between the meshes. Thus, a single template mesh is selected to define initial correspondences, and all other meshes are registered and sampled at vertices with the closest distance to the template. Consequently, all meshes share the same number of vertices where anatomical landmarks are roughly aligned. For efficiency, the meshes are down-sampled. In this work, original patient meshes are used, but in general, their corresponding vertices can also be used to build a statistical shape model that allows the generation of arbitrary meshes. 

\subsection{Data generation}
\label{subsec:methods:datageneration}
A 2D US image displays a cutplane through the heart depending on the 3D pose of the US probe. The pipeline introduced by Gilbert et al.  \cite{gilbert2021} was extended to generate ground truth annotations. The aim is to sample 2D images from the 3D mesh that mimic realistic views obtained by a cardiologist. Alignment of the model towards specific standard views can be achieved using anatomical landmarks. For example, the apical 4 chamber view (a4ch) cuts the heart along a virtual axis built between the LV apex perpendicular to the midpoint between the mitral and tricuspid valve. Note that this only serves as a reference and can be highly patient-specific in clinical practice. 
After defining those standard planes, arbitrary other planes can be sampled by varying scaling, rotation, and translation of the 3D mesh which is aligned to the coordinate system built by the landmarks, e.g., the a4ch view origin (0,0,0) is set to the intersection between the mitral valve and tricuspid valve. With this, planes reflect the natural acquisition variability, and a cut through the mesh results in a 2D multi-label segmentation. %It must be noted here that the original fine-grained meshes are used for the segmentation generation. 
Segments are further concatenated with randomly sampled black cones to mimic the sector around a typical US image. For further details on the pipeline, the reader is referred to \cite{gilbert2021}. Unlike the original pipeline, the output mesh is defined in the plane coordinate system; \textit{x} and \textit{y} coordinates are translations on the image plane, and \textit{z} represents the depth. In addition to generated segmentation, 3D vertices and the closest standard view are stored.

%- Project unstructured tetrahedral grid and polydata on plane
%- augmentation is done by 3D variations (translate,rotate,scale)
%%- save points (x,y,z) in image coordinate system
%- normalize points to -1 and 1 
%- save best estimate for plane
%- save mask 
%- save pseudo images

\subsection{Diffusion model data generation}

Different echocardiography datasets have been published \cite{Leclerc2019a,Ouyang,cetus2014}. However, no dataset with all the four chambers is available, which would be required to train the GCN in a supervised manner. Labeling 3D US images is time-consuming and challenging due to their varying quality. Instead, a diffusion model is used to generate realistic US images based on the segmentations from sec. \ref{subsec:methods:datageneration}. Diffusion models~\cite{DBLP:journals/corr/abs-2006-11239} have shown promising results in generating realistic and diverse images, including medical images. These models iteratively transform sampled noise into a more complex, realistic image. The training involves a forward path with multiple diffusion steps, where noise is increasingly added to the training image, and a reverse path to gradually denoise the noise source to generate a sample. Since the diffusion process aims to model the gradual transition from the initial noise source to the target image distribution, these models generate high-quality samples that progressively become more realistic. Building on the research of Tiago et al.~\cite{tiago2023}, an adversarial denoising diffusion model is combined with a generative adversarial network (GAN) to learn the reverse denoising process, which combines the advantages of two generative models. The GAN attempts to create realistic image samples, whose statistical distribution is similar to the original, utilizing an extra label, particularly a segmentation mask, to simplify the image generation process. Using the segmentation masks described in sec. \ref{subsec:methods:datageneration}, the goal is for the generated images to look like clinical images and align with the cardiac structures, so that they can be used as training data for the graph neural network.

\subsection{Graph neural network}
\label{subsec:methods:gcn}
A 3D mesh can be defined by a graph $G = (V,E)$ in which nodes $V =\{v_i | i =1, ...,N\}$ represent vertices and $E$ represent edges of the surface faces. Unlike CNNs, graph neural networks (GCNs) allow for operating on arbitrary non-Euclidean structured data by aggregating the information of neighboring nodes using the edges and a weighting term. Spatial changes of a vertex likely affect neighboring vertices, but may also affect other vertices that are not in direct spatial proximity. There are different ways to define edge relations in an adjacency matrix. Gong et al. \cite{gong2019spiralnet} propose using spiral convolutions on a closed surface. This operator enforces a fixed ordering of neighboring nodes during message passing to compute node updates with $x^k_i = \gamma^k \left( \parallel_{j\in S(i,l)}  x_j^{k-1} \right)$ where $S(i,l)$ is the fixed spiral concatenation of indices of neighboring vertices $x_j$ and $\gamma$ is a multi-layer perceptron.
Here, we extend the work in \cite{gcn2022miccai} to model a 3D heart mesh. An adjacency matrix is built for each structure and is concatenated. Similarly to the 2D approach, each graph node is assigned with a feature vector as the initial input. For this purpose, a CNN (ResNet50 \cite{he2016deep}) predicts a feature representation from the input image. The GCN decoder is equipped with an initial dense layer to compress the input features, followed by four spiral layers, and complemented by an exponential linear activation unit. The last GCN layer outputs a 3D vector for each node and represents a mesh aligned with the image. The network is optimized using an L2 loss between the predicted and ground truth vertices. 

%\todo{here further experiments are already implemented with different loss functions}%We separate the x,y and the z coordinates to assign different weights during optimization since in-plane coordinates are regarded to be more important. %Additionally, the view label can guide the optimization process especially when differentiating between adjacent views. Experimental studies showed that this loss only helps after warming up the network. Additionally, two more losses are integrated in this fine-tuning step. We compute a distance map of the labels so that we can put more emphasis on how the model performs at surface points that are close to the image plane which ideally should align with the label borders. In addition, another weighting term prioritize points closer to the image plane.

\section{Experiments \& Results}
\label{sec:experiments}

%Implementation details
%spiral length 
%dilation
% hyperparameter

The proposed GCN generates a 3D mesh from an US image. The performance can be evaluated on multiple downstream tasks important for clinical workflow optimization, such as segmentation of anatomical structures, view recognition, and foreshortening detection. In this proof of concept, experiments focused on view recognition and structure localization to answer the following questions:
\begin{itemize}
    \item [Q1] Can the GCN be used to predict a 3D representation from a 2D image? %(sec. \ref{subsubsec:experiments:reco})
    \item [Q2] Trained solely on synthetic data, can the GCN predict a view label from a synthetic and a clinical US image? %(sec. \ref{subsubsec:experiments:viewrec})
    \item [Q3] Can the GCN accurately localize the four different chambers in the image? %(sec. \ref{subsubsec:experiments:detection})
\end{itemize}

\subsubsection{Data and Implementation:}
\label{subsubsec:experiments:data}
Different data sources were used for the experiments. 4258 synthetic segmentations were sampled from 20 patient meshes processed by the data generation pipeline following \cite{gilbert2021}. All meshes were downsampled to a total of 2008 vertices. Based on the four standard views, 4-chamber (\textit{a4ch}), 2-chamber (\textit{a2ch}), 5-chamber (\textit{a5ch}), and apical long axis (\textit{aplax}), variations of the target planes were augmented by applying rotation, translation and scaling parameters sampled from a uniform distribution. Distribution limits were chosen for each view to reflect clinical variation. For training the diffusion model, an internal 2D echocardiography dataset was used since there was no public dataset with 4 chambers. This dataset consists of 1318 training and 248 test images from multiple sites and US probes and the aforementioned four standard views for which two timesteps, namely end-diastole (ED) and end-systole (ES), were labeled by experienced cardiologists. Labeled structures include ventricles \textit{LV} and \textit{RV} and atriums \textit{LA} and \textit{RA}. The diffusion model was applied to the synthetic segmentations to create 4258 synthetic US images. All datasets were divided into training, validation and test. Different appearance-based augmentations (CLAHE, gamma, multiplicative and Gaussian noise) were applied. The network was trained with a batchsize of 32 and optimized with a learning rate of 4e-4 using ADAM~\cite{kingma2014adam} until convergence.
%\todo{scanned with a GE scanner, which model and add sampling parameters for the model,GPU size}
% Add more details here is requested by the review, not enough space to do it now

\subsubsection{View recognition:}
\label{subsubsec:experiments:viewrec}
\begin{figure}[t!]
    \centering
    \includegraphics[width=0.94\textwidth]{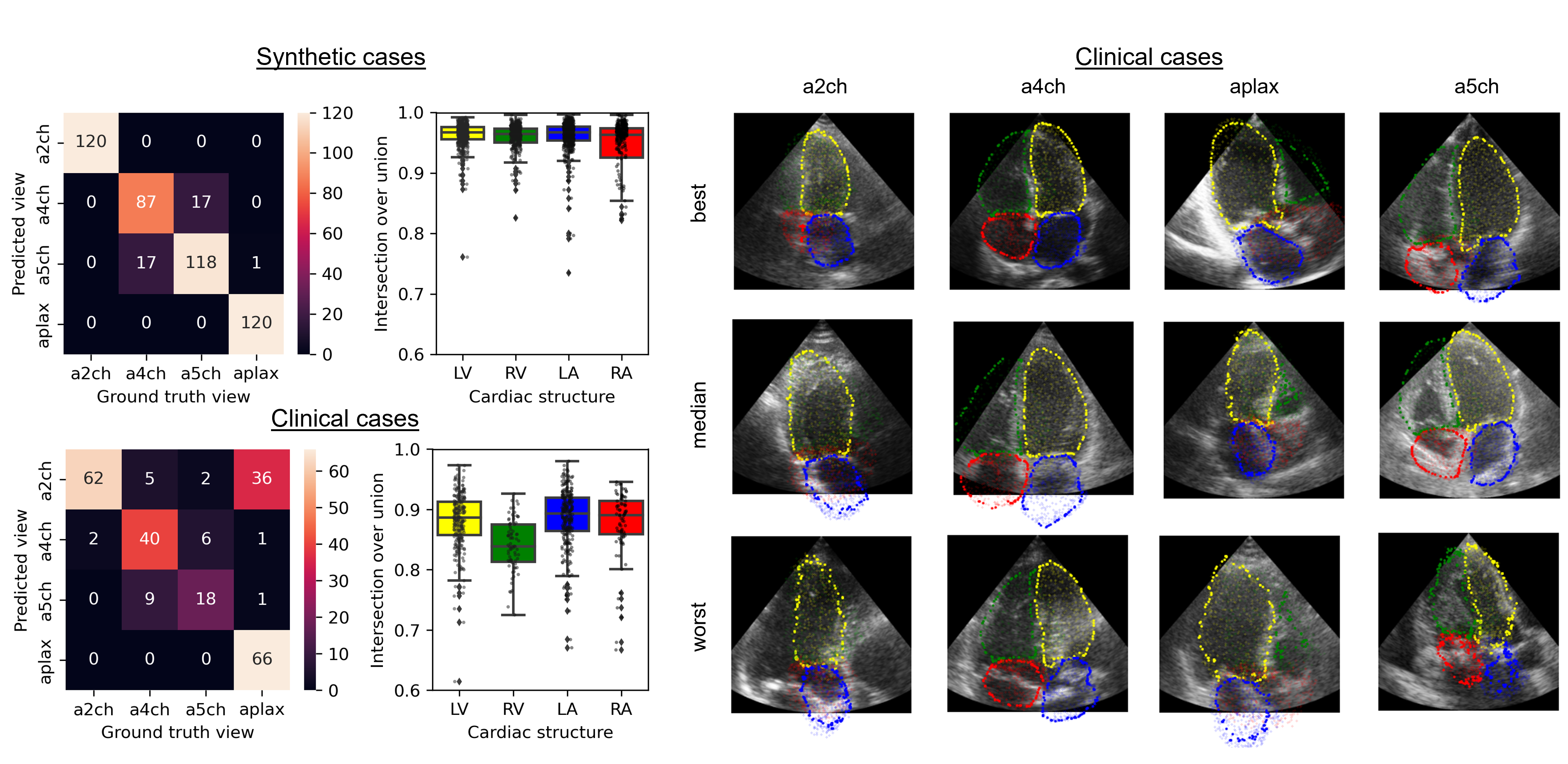}
    \caption{Left: Confusion matrix of the view prediction and bounding box IoU for 480 synthetic and 248 clinical test cases. Right: Qualitative examples of the projected 3D GCN output applied to clinical cases (best, median, worst based on mIoU). The larger dots indicate where the mesh intersects the image plane (see example in suppl. video).}
    \label{fig:visualexamples}
\end{figure}

Four standard views are encoded in the template mesh by computing the cutplane and identifying all vertices close to that plane. Those corresponding vertices can be used to transfer the ground truth plane to the predicted mesh and compare it to the plane spanned by the predicted vertices that intersect with the image plane. Results are shown in Table~\ref{tab:resultsview} and Fig.~\ref{fig:visualexamples}. In most synthetic test cases, the model successfully derived the correct label. However, when applied to clinical cases, the model struggled to differentiate between \textit{a5ch} and \textit{a4ch} and between \textit{aplax} and \textit{a2ch}. To benchmark this performance drop, a ResNet classifier was trained as a reference in addition to the 3DGCN using the same pipeline. The implementation included a ResNet50 backbone pre-trained on ImageNet with similar data and augmentation. The 3DGCN was able to outperform the ResNet, although it should be noted that training conditions were not specifically targeted towards the ResNet. The results show that the domain gap between synthetic and clinical images is present for the ResNet, with an even larger performance drop. The observed domain gap could be attributed to differences in how chambers are labeled in clinical images. In particular the LVOT, which was not included in the clinical annotations, is among the main distinguishable features of the \textit{a5ch} and \textit{aplax} view. This might have adversely affected the synthetic image generation that in return lead to inaccuracies in the view prediction. Regardless of the model, distinguishing between \textit{a4ch} and \textit{a5ch} views seems to be most challenging as those two only differ by a tilt of approx. 10-20\textdegree\ around the LV apex, which can be difficult to discern.

\begin{table}[t]
\centering
\begin{tabular}{@{}p{1.4cm}lclllllllll@{}}
\toprule
        Method & \multicolumn{2}{l}{Data}
        & \hspace{3mm} weighted avg
        &\multicolumn{2}{l}{a2ch} 
        &\multicolumn{2}{l}{aplax} 
        &\multicolumn{2}{l}{a5ch} 
        &\multicolumn{2}{l}{a4ch} \\
  &Train  & Test &\hspace{3mm} view acc. \hspace{1cm}&
  prec. & recall& 
  prec. & recall& 
  prec. & recall& 
  prec. & recall\\
\midrule
%ResNet  & segm    & segm       &  -  &  -   & 0   \\ 

ResNet50  & syn   & syn&\hspace{3mm} 0.88 \hspace{8mm}& 1.00 & 1.00 & 0.60 & 0.78 & 0.88 & 0.76 & 1.00 & 0.99  \\ 
ResNet50  & syn    & real&\hspace{3mm} 0.46\hspace{8mm} & 0.89 & 0.53 & 0.02 & 1.00 & 0.92 & 0.19 & 0.12 & 0.87  \\ 
3DGCN  & syn   & syn&\hspace{3mm} 0.93\hspace{8mm} & 1.00 & 1.00 & 0.99 & 1.00 & 0.87 & 0.87 & 0.84 & 0.84  \\ 
3DGCN  & syn  & real &\hspace{3mm} 0.75\hspace{8mm} & 0.97 & 0.59 & 0.63 & 1.00 & 0.69 & 0.64 & 0.74 & 0.82 \\ 
\midrule
\end{tabular}
\caption{Quantitative results for the task of view recognition evaluated on 480 synthetic and 248 clinical test cases. 'syn' refers to the images created by the diffusion model and 'real' refers to clinical cases. The method is compared to a standard classification network (ResNet50) based on accuracy, precision, and recall.}

\label{tab:resultsview}
\end{table}

\begin{table}[t]
\centering
\begin{tabular}{@{}p{1.5cm}lcllllc@{}}
\toprule
        Method & \multicolumn{2}{l}{\hspace{0.5cm}Data} &\multicolumn{4}{c}{mIoU} &\hspace{1cm} {mkptsErr [\%]}\hspace{1cm} \\
  & Train  & Test\hspace{1cm} & LV & LA & RV & RA & \\ 

\midrule
3DGCN  & segm    & segm\hspace{1cm}   & 0.97 & 0.96& 0.93& 0.83  &  $2.16 \pm0.74$      \\ 
3DGCN  & syn    & syn\hspace{1cm} & 0.96 & 0.96 & 0.89 & 0.79 & $2.68 \pm1.13$   \\ 
3DGCN  & syn  & real\hspace{1cm} & 0.88 & 0.88& 0.82& 0.87  & N/A     \\ 
\midrule
\end{tabular}
\caption{Quantitative results for the task of structure localization evaluated on 480 synthetic and 248 clinical test cases based on the bounding box mIoU of different structures and the mean 3D keypoints error. 'segm' refers to the segmentations, 'syn' refers to the images created by the diffusion model and 'real' refers to clinical cases.}

\label{tab:resultsbb}
\end{table}

\subsubsection{2D/3D Reconstruction:}
\label{subsubsec:experiments:reco}
This experiment can only be evaluated on the synthetic and segmentation datasets as no ground truth coordinates are available for clinical cases. Here, \textit{segmentation} refers to the original segmentations, which should be a simple task for the graph. The mean 3D coordinate error (mkptsErr) is measured between the ground truth and the prediction and averaged over all 3D points. Since synthetic images do not have absolute pixel values, we computed the error in relation to the image size. As shown in Table \ref{tab:resultsbb}, the kpts error is higher for synthetic US images, likely caused by the distribution shift between the segmentation masks and the synthetic data created by the diffusion model, as the latter could introduce unnatural appearance variations or inaccuracies.

\subsubsection{Structure localization:}
\label{subsubsec:experiments:detection}
Intersecting the 3D mesh with the image plane leads to multiple 2D contours that allow semantic segmentation. However, a direct comparison is unfeasible due to structural differences between synthetic and clinical segmentations. Instead, bounding boxes are analyzed, and the mean Intersection over Union (mIoU) is computed at box level rather than pixel level. For the bounding box, points are considered that lie closer than 5\% of the image size. The results presented in Fig. \ref{fig:visualexamples} and Table \ref{tab:resultsbb} demonstrate good performance on synthetic data but also a noticeable performance drop on clinical data.

%\vspace{-0.8cm}
\section{Discussion \& Conclusion}
\label{sec:discussion}
%\todo{Add SOTA models like ResNet/VGG net and detection net in experiments?}
We present a novel approach solving cardiac view recognition. Based on a US image, a GCN reconstructs a 3D mesh to extract the contours of four chambers and to estimate the pose and view. To tackle the lack of 3D ground truth annotations, a diffusion model was employed to generate realistic US images based on binary masks. The GCN was solely trained on those images demonstrating promising potential for both view recognition and structure detection when applied to synthetic cases. The results indicate a domain gap in the performance when testing on synthetic versus clinical data, possibly attributed to the different ways of labeling clinical cases --- particularly the lack of the LVOT. Furthermore, the diffusion model was trained on ED and ES frames, but the 3D meshes only stem from the ED phase. This might limit the performance on the full heart cycle and may be solved by incorporating a 4D model. Some failure cases could be attributed to low image quality and unusual heart shapes. A diffusion model trained on limited data can only create an approximation of the target distribution. Supplementing synthetic with real clinical 3D US images could help to ultimately close the domain gap. This study aims to showcase a holistic approach to view recognition, which can facilitate other tasks like chamber segmentation and pose estimation. Chamber localization was only an auxiliary task to compare the performance between real and synthetic images. Future experiments will assess the segmentation quality directly as soon as the LVOT is annotated. Still, chamber localization can serve as a feature for automatic image quality control. With the resulting 3D mesh, an intersection of structures with the US sector and foreshortening could potentially be detected which needs to be verified in the future. Since the 3D mesh includes more anatomical structures, their relative location encoded in the mesh can also be estimated as opposed to semantic segmentation. The 3DGCN can handle occlusions (illustrated in Fig. \ref{fig:visualexamples}), both within and outside the image plane and can be trained on multiple views, eliminating the need for view-specific keypoints, which are required for the 2DGCN \cite{gcn2022miccai}. In the future, the GCN can be extended to fuse several views to refine the 3D mesh further. Using the original patient meshes from \cite{rodero_cristobal_2021_4590294}, we aimed to provide the diffusion model with shapes close to the training distribution. In the future, we plan to add a statistical shape model as described in \cite{rodero_cristobal_2021_4506930}.
While the full potential has not yet been fully demonstrated in clinical cases, this approach stands out from other black-box view recognition methods by providing direct visual feedback, thereby increasing explainability in failure cases. In conclusion, the concept of applying GCNs to view recognition seems to be promising as it combines different tasks relevant to assist the cardiologist in quality control.

\textbf{Acknowledgement:}
We thank Anna Novikova and Daria Kulikova for their valuable clinical consultation and for annotating the training data.
\bibliographystyle{splncs04}
\bibliography{paperXXXX}

%\newpage

%\section{Supplementary material}
%\begin{figure*}[htbp]
%    \centering
%    \includegraphics[width=\textwidth]{synthetic_cases.png}
%    \caption{Left: Qualitative examples of the projected 3D graph output from the GCN applied on synthetic cases (best,median,worst).}
%    \label{fig:visualexamples2}
%\end{figure*}
\end{document}

\title{Supplementary material: Graph Convolutional Neural Networks for
Automated Echocardiography View Recognition: A Holistic Approach}
\author{}
\institute{}
\maketitle
\vspace{-15mm}
\section{Training and implementation details}
\vspace{-6mm}
%\subsection{Training parameters GCN:}
\begin{table}[h!]
\centering
\begin{tabular}{@{}p{3.5cm}p{9cm}@{}}
Parameter & Value \\ \hline \\
 Training data &  3600 synthetic images created from 16 patient meshes \cite{rodero_cristobal_2021_4590294} \\
 Validation data &  478 synthetic images created from 2 patient meshes \cite{rodero_cristobal_2021_4590294}\\
 Test data &  Syn: 480 synthetic images created from 2 patient meshes \cite{rodero_cristobal_2021_4590294}\\
 & Real: Internal dataset, GE scanner, 248 images \\
 Backbone & ResNet 50 \\
 GCN layers & [4, 8, 8, 16, 16, 32, 32, 48] \\
Input image size & [1,256,256] \\ 
Batch size & 32 \\
Number of epochs & 400 \\
 Optimization & Adam \cite{kingma2014adam} with a learning rate of 4e-4\\ 
 Pre-training & ImageNet \\ 
 Augmentation & python package albumentations\cite{2018arXiv180906839B}\\
 & Random Gamma (default)\\
 & RandomBrightnessContrast (default) \\
 & Multiplicative Noise with multiplier=(0.9, 1.1) \\
 & Gaussian Noise with (10.0,50.0) \\
 & CLAHE, clip limit=6.0, tile grid size=(10, 10)\\
 Training time &  ca. 2 days (1 Nvidia Quadro P6000) \\

\end{tabular}
\label{tab:networks}

\caption{Details on data, hyperparameters and applied augmentations of the 3DGCN}
\end{table}

%\subsection{Training parameters DDM:}
\begin{table}[h!]
\centering
\begin{tabular}{@{}p{3.5cm}p{9cm}@{}}
Parameter & Value \\ \hline \\
 Training data &  Internal dataset, GE scanner, 1040 images\\
 Validation data &  Internal dataset, GE scanner, 278 images\\
 Test data &  4258 synthetic images created from 20 patient meshes \cite{rodero_cristobal_2021_4590294} \\
 Input image size & [1,256,256] \\ 
 Batch size & 2 \\
 Number of epochs & 1000 \\
 Diffusion steps & 4 \\
 Optimization & Adam \cite{kingma2014adam} with a learning rate of 1e-4\\ 
Training time & ca. 1 week (2 Nvidia RTX 2080 ) \\
\end{tabular}
\label{tab:networks}

\caption{Details on data and hyperparameters of the DDM}
\end{table}

\section{Visual training examples - synthetic data}
  \vspace{-0.7cm}
\begin{figure}[]
    \centering
    \includegraphics[width=0.8\textwidth]{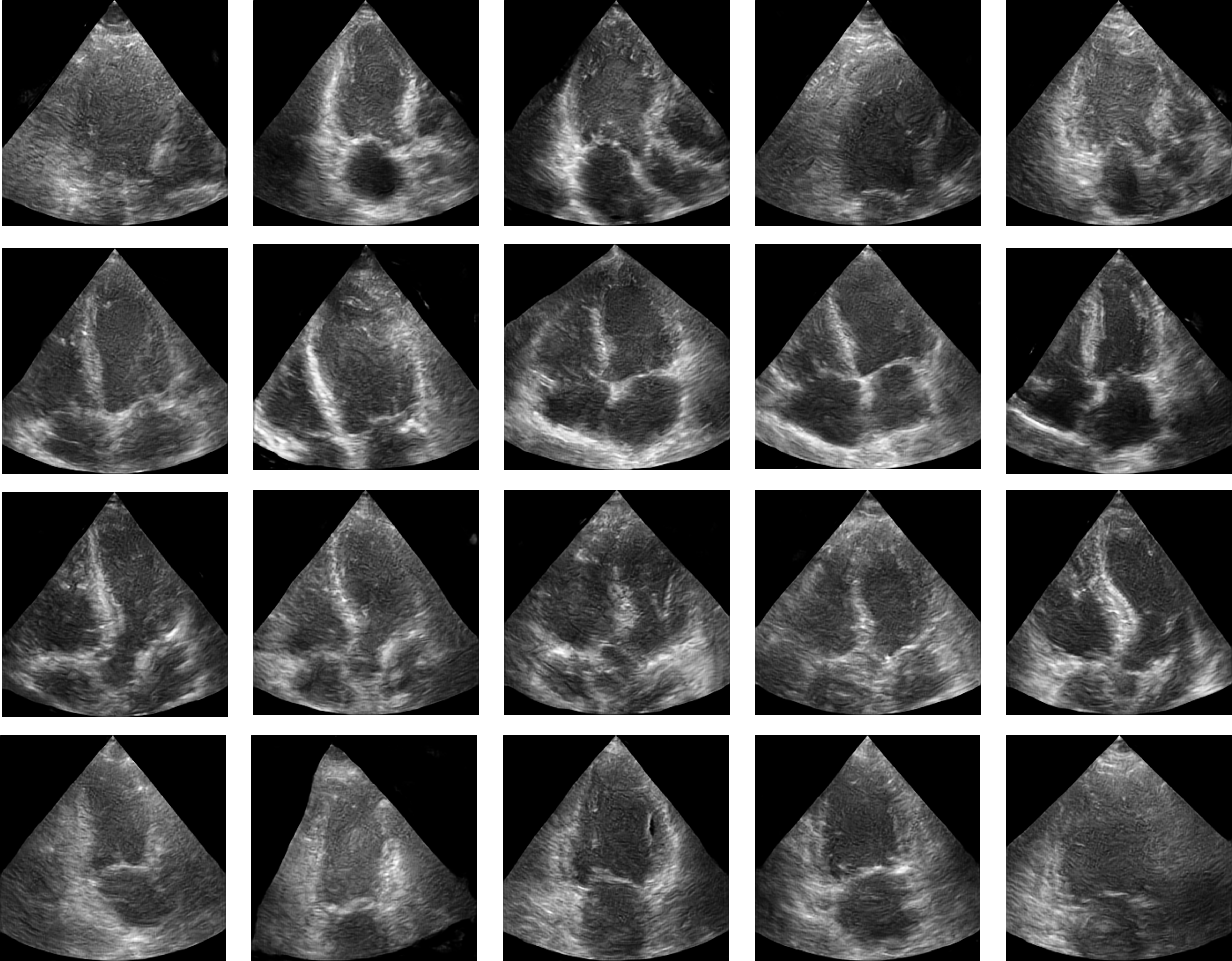}
    \vspace{-0.2cm}
    \caption{ Training examples to show case the output of the diffusion model. From top to bottom: 1. aplax 2. a4ch 3. a5ch 4. a2ch } 
    \label{fig:visualexamples2}
\end{figure}
  \vspace{-1cm}
\section{Annotations}
  \vspace{-0.7cm}
\begin{figure}[]
    \centering
    \includegraphics[width=0.45\textwidth]{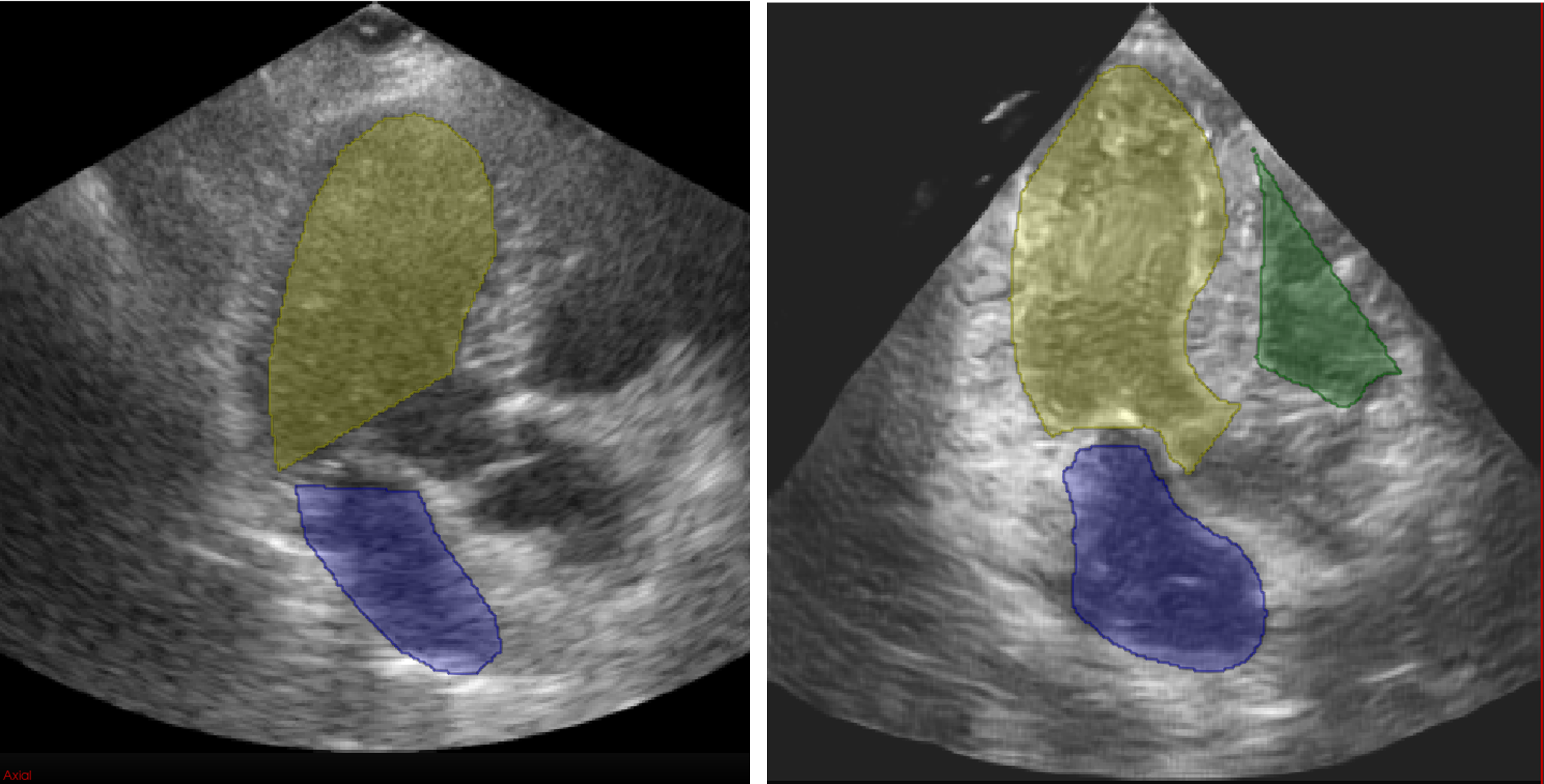}
    \vspace{-0.2cm}
    \caption{Difference between synthetic and real images. Left: Real aplax image with coarse manual segmentation without LVOT. Right: Synthetic image created by the diffusion model with segmentation sampled from the 3D mesh and LVOT.} 
    \label{fig:visualexamples2}
\end{figure}
\newpage
\vspace{-0.9cm}
\section{Visual test examples - synthetic data}
\begin{figure}[]
    \centering
    \includegraphics[width=\textwidth]{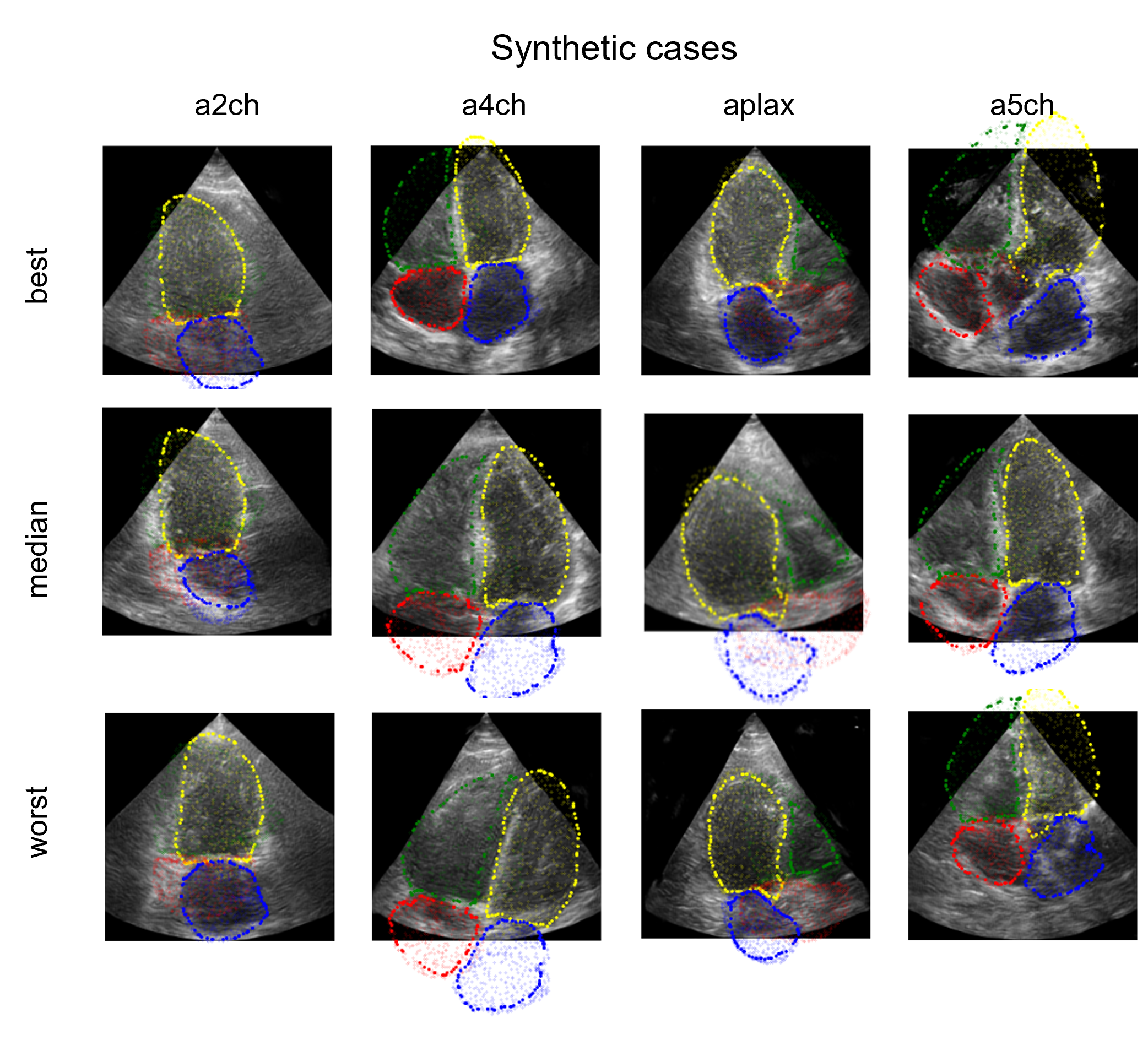}
    \vspace{-0.2cm}
    \caption{ Qualitative examples of the projected 3D graph output from the GCN applied on synthetic cases (best,median,worst based on mIoU)} 
    \label{fig:visualexamples2}
\end{figure}

\vspace{-0.9cm}

\bibliographystyle{splncs04}
\bibliography{paperXXXX}